\documentclass[preprint2]{proto}
\usepackage{times}
\newcommand{\refs}{\par\noindent\hangindent=1pc\hangafter=1}
\newcommand{\kms}{km\thinspace s$^{\rm -1}$}

\newcommand{\ts}{\thinspace}

\newcommand{\msun}{M$_{\odot}$}

\begin{document}

\title{\textbf{\LARGE Toward Resolving the Outflow Engine: An Observational
Perspective}}

\author{\textbf{\large Tom Ray}}
\affil{\small\em Dublin Institute for Advanced Studies}
\author{\textbf{\large Catherine Dougados}}
\affil{\small\em Laboratoire d'Astrophysique de Grenoble}
\author{\textbf{\large Francesca Bacciotti}}
\affil{\small\em INAF-Osservatorio Astrofisico di Arcetri}
\author{\textbf{\large Jochen Eisl\"offel}}
\affil{\small\em Th\"uringer Landessternwarte Tautenburg}
\author{\textbf{\large Antonio Chrysostomou}}
\affil{\small\em University of Hertfordshire}

\begin{abstract}
\baselineskip = 11pt
\leftskip = 0.65in 
\rightskip = 0.65in
\parindent=1pc

{\small 
Jets from young stars represent one of the most striking signposts of star
formation. The phenomenon has been researched for over two decades 
and there is now general agreement that such jets are generated as a by-product 
of accretion; most likely by the accretion disk itself. Thus they mimic 
what occurs in more exotic objects such as active galactic nuclei and 
micro-quasars. The precise mechanism for their production however remains 
a mystery. To a large degree, progress is hampered observationally by the 
embedded nature of many jet sources as well as a lack of spatial resolution: 
Crude estimates, as well as more sophisticated models, nevertheless suggest that 
jets are accelerated and focused on scales of a few AU at most. 

It is only in the past few years however that we have begun to probe such 
scales in detail using classical T Tauri stars as touchstones. Application 
of adaptive optics, data provided by the HST, use of 
specialised techniques 
such as spectro-astrometry, and the development of spectral diagnostic tools, 
are beginning to reveal conditions in the jet launch zone. This has helped 
enormously to constrain models. Further improvements in the quality of 
the observational data are expected when the new generation of interferometers 
come on-line. Here we review some of the most dramatic findings in this 
area since Protostars and Planets~IV including indications for jet 
rotation, i.e.\ that they transport
angular momentum. We will also show how measurements, such as those of 
width and the velocity field close to the source, suggest jets are initially
launched as warm magneto-centrifugal disk winds.  

Finally the power of the spectro-astrometric technique, as a probe of the 
central engine in very low mass stars and brown dwarfs, is shown by revealing 
the presence of a collimated outflow from a brown dwarf for the first time, 
copying what occurs on a larger scale in T~Tauri stars.
\\~\\~\\~} 
 
\end{abstract}  

\section{\textbf{INTRODUCTION}}

The phenomenon of jets from young stellar objects (YSOs) has been known for 
over two decades. While we now have a reasonably good understanding of how they 
propagate and interact with their surroundings on large scales (e.g.,  
see the chapter by {\em Bally, Reipurth and Davis}), i.e.\ hundreds of AU and 
beyond, how precisely these jets are generated remains a puzzle. The 
observed correlation between mass outflow and accretion through the star's disk 
(e.g.\ {\em Hartigan et al.}, 1995; {\em Cabrit et al.}, 1990) would seem to 
favour some sort of 
magnetohydrodynamic (MHD) jet launching mechanism but which one is open to 
question. In particular it is not known whether jets originate from the 
interface between the star's magnetosphere and disk (the so-called X-wind 
model, see the chapter by {\em Shang et al.}; {\em Shang et al.}, 2002; {\em 
Shu et al.}, 2000) or from a wide range of disk radii (the 
disk or D-wind model, see the chapter by {\em Pudritz et al.}).  

\begin{figure*}[ht]
\centering
\includegraphics[width=15.9cm]{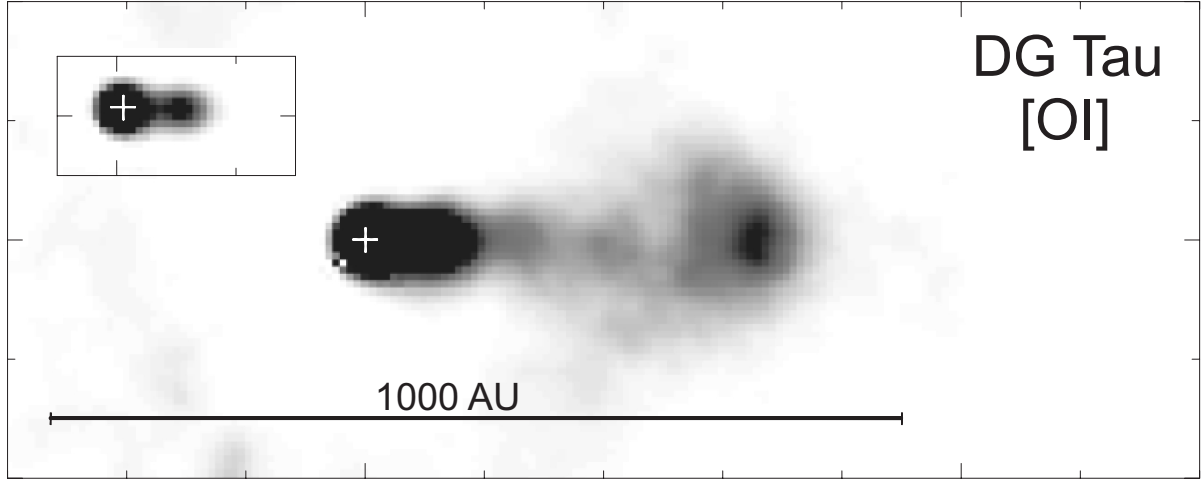}
\caption{Deconvolved [OI]$\lambda$6300 + continuum narrow-band image of the 
DG~Tau jet obtained with the AO system PUEO on the Canada-France-Hawaii 
Telescope. The spatial resolution achieved was 0\farcs 1 = 14 AU 
at the distance of the Taurus Auriga Cloud. Inset (top-left) is a high 
contrast image near the source. Adapted from {\em Dougados et al.,} (2002).}
\label{dgoi}
\end{figure*}

On the observational front we have begun to probe the region where the 
jet is generated and collimated, thus testing the various models. Moreover 
since Protostars and Planets IV a number of major advances have been made 
thanks to the availability of high angular resolution imaging and spectroscopy. 
In particular the use of intermediate dispersion spectroscopy 
(long-slit, Fabry-Perot or employing integral field units) has provided 
excellent contrast between the line emitting outflow and its 
continuum-generating parent YSO, a pre-requisite to trace outflows right back 
to their source (see Section 2). Examples include ground-based 
telescopes equipped with Adaptive Optics (AO) and the Space Telescope Imaging 
Spectrograph (STIS), giving angular resolution down to 0\farcs 05 
(see Section 2 \& Section 3). 
In addition, as will be explained below, the technique of spectro-astrometry
(see Section 5) is providing insights on scales of a few AU from the source. 
The new methodologies have brought an impressive wealth of morphological 
and kinematical data on the jet-launching region ($\le$~200~AU from the YSO) 
providing the most stringent constraints to date for the various models. 
Here we will illustrate these results, e.g.\ measurements of the jet diameter 
close to the source, and, where appropriate, comparison with model predictions. 

Jets may be Nature's way of removing excess angular momentum from accretion 
disks, thereby allowing accretion to occur. Moreover, they are produced not 
only by young stars but by a plethora of astronomical objects from nascent 
brown dwarfs, with masses of around 5.10$^{\rm -2}$\msun, 
to black holes at the centre of AGN, as massive as 5.10$^{\rm 8}$\msun. 
In between they are generated by X-ray binaries, symbiotic systems, planetary 
nebulae, and gamma ray burst sources (e.g., {\em Livio}, 2004). The range of 
environments, and the astounding 10 orders of magnitude in mass over which the 
jet mechanism operates, is testimony to its robustness. Thus understanding how 
they are generated is of wide interest to the Astrophysics Community. Given 
the quality of data now coming on stream, and the prospect of 
even better angular resolution in the near future, we are hopefully close to 
unravelling the nature of the engine itself. Our suspicion, as supported below,
is that this will be done first in the context of YSO outflows.     

\section{\textbf{IMAGING STRUCTURES CLOSE TO THE JET BASE}}

YSO jets largely emit in a number of atomic and molecular 
lines (see, for example, {\em Reipurth and Bally}, 2001 or 
{\em Eisl\"offel et al.}, 2000). In the infrared to
the ultra-violet, these lines originate in the radiative cooling zones of 
shocks with typical velocities from a few tens to a few hundred \kms . Thus 
in order to explore the morphology 
and kinematics of the jet launching zone, both high spatial resolution 
narrow-band imaging (on sub-arcsecond scales) and  intermediate resolution 
spectroscopy is required. Leaving aside interferometry of their meagre radio 
continuum emission ({\em Girart et al.}, 2002), currently the best spatial 
resolution is afforded by optical/NIR instruments. Many YSOs are so 
deeply embedded however
that optical/NIR observations are impossible. That said, one class of evolved 
YSO, namely the classical T Tauri stars (CTTS), are optically visible and have 
jets (see Fig.~\ref{dgoi}). CTTS therefore offer 
a unique opportunity to test current ejection theories. In particular, they
give access to the innermost regions of the wind ($\leq$ 100 AU) where models 
predict that most of the collimation and acceleration processes occur. 
Moreover the separate stellar and accretion disk properties of these systems 
are well known. Here we summarise what has been learned from recent 
high-angular 
resolution imaging studies of a number of CTTS jets. These observations 
have been conducted either from the ground with adaptive optics (AO) 
or from space with the Hubble Space Telescope (HST).

\begin{figure}[ht]
\centering
\includegraphics[width=8cm]{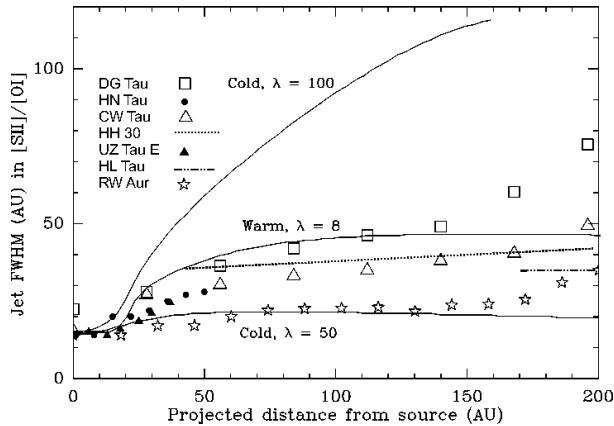}
\caption{Variation of jet width (FWHM) derived from [SII] and [OI] images as a 
function of distance from the source. Data points are from 
CFHT/PUEO and HST/STIS observations made by the authors as well as those of 
{\em Hartigan et al.,} (2004). Overlaid (solid lines) are 
predicted variations based on two cold disk wind models with low 
efficiency (high $\lambda$) and a warm disk solution for 
comparison from {\em Dougados et al.,} 
(2004). Note that moderate to high efficiency is favoured, i.e.\ warm 
solutions. Here efficiency is measured in terms of the ratio of mass 
outflow to mass accretion. Models are convolved with a 14~AU (FWHM)
gaussian beam. For full details see {\em Garcia et al.,} (2001) and 
{\em Dougados et al.,} (2004).}
\label{widths}
\end{figure}

A fundamental difficulty in imaging a faint jet close to a bright 
CTTS is contrast with the 
source itself. This problem is often further exacerbated by the presence of an 
extended reflection nebula.  Contrast with the line emitting jet 
can be improved either by decreasing the PSF (with AO systems from 
the ground or by imaging from space with, e.g., HST) and by increasing the
spectral resolution. Obviously the optimum solution is a combination
of both.  The pioneering work of {\em Hirth et al.,} (1997) used
long-slit spectroscopy with accurate central continuum subtraction to
reveal spatial extensions of a few arcseconds in the forbidden line
emission of a dozen CTTS.  Around the same time, narrow-band imaging with the
HST provided the first high spatial resolution images of jets from CTTS 
({\em Ray et al.}, 1996). More recently application of AO systems from the 
ground on 4m-class telescopes, using both conventional narrow-band imaging and 
in combination with intermediate spectral resolution systems, 
have led to remarkable 0\farcs 2 resolution images  
of a number of small-scale jets from CTTS including DG~Tau, CW~Tau, and 
RW~Aur ({\em Dougados et al.}, 2000; {\em Lavalley et al.}, 1997; 
{\em Lavalley-Fouquet et al.}, 2000). 
An alternative approach, pioneered by {\em Hartigan et al.,} (2004), is 
$``$slit-less spectroscopy$''$ using high spatial resolution instruments such as
the STIS on board the HST. The advantage of this method
is that, providing the dispersion of the spectrograph is 
large enough, one can obtain non-overlapping emission-line images covering a 
large range of wavelengths. This includes lines where no narrow-band HST 
filters exist. In this way {\em Hartigan et al.,} (2004) constructed high 
spatial resolution images 
of the jets from CW~Tau, HN~Tau, UZ~Tau~E, DF~Tau, and the primary of DD~Tau. 
Moreover, as will be described fully in the next section, the use of 
contiguous parallel long-slits, for example with STIS, can also provide high 
resolution ``images'' 
not only in individual lines but over a range of velocities 
(see Section 3 and, for example, {\em Woitas et al.}, 2002). 

All of these observations show that these $``$small-scale$''$ T~Tauri jets
have complex morphologies, dominated by emission knots, that strongly
resemble the striking HH jets emanating from IRAS Class~0/1 sources (see the 
chapter by {\em Bally, Reipurth, and Davis}). Turning to larger scales, 
deep ground-based imaging by {\em Mundt and Eisl\"offel,} 
(2000) revealed that CTTS jets are associated with 
faint bow-shock like structures, at distances of a few thousand AU. 
Moreover {\em McGroarty and Ray}, (2004) found that CTTS outflows 
can stretch for several parsecs as is the case with outflows from 
more embedded YSOs. 
Thus the  observations clearly suggest that the same ejection
mechanism is at work at all phases of star formation. 

Images taken at different epochs, usually a few years apart, 
can reveal how the outflow varies with time. For example, large proper 
motions, on the order of the jet flow velocity, have been inferred for 
the knots in the DG~Tau and RW~Aur jets ({\em Dougados et al.}, 2000; {\em 
L\'opez-Mart\'{\i}n et al.}, 2003; {\em Hartigan et al.}, 2004). In the DG~Tau 
jet, a clear bow-shaped morphology is revealed for the knot located at
3$^{\prime\prime}$ ({\em Lavalley et al.}, 1997; {\em Dougados et al.},
2000) as well as jet wiggling, suggestive of precession. These
properties, as in the younger HH~flows, suggest that knots are likely to 
be internal working surfaces due to time variable ejection.  Detailed
hydrodynamical modelling, for example of the DG~Tau jet, indicate 
fluctuations on time scales of 1-10 yrs, although one single period 
does not seem to account for both the kinematics and the morphology 
({\em Raga et al.}, 2001). Such rapid changes are also observed at the base of 
the younger, embedded HH~flows. Attempts at modelling these changes have 
been made. For example, numerical simulations by {\em Goodson et al.,} 
(1999) of the interaction of the stellar magnetosphere with the disk 
predict a cyclic inflation of the former leading to eruptive ejection. 
The estimated timescales, however, seem too 
short as they are around a few stellar rotation periods. 
It is also possible that such short term fluctuations in the outflow 
may come from disk instabilities or cyclic changes in the stellar 
magnetic field (as in the Sun). Further study is clearly needed to 
give insight into the origin of the variability process.

Imaging studies close to the YSO reveal details that are useful in 
discriminating between various models. For example
plots of jet width (FWHM) against distance from the source, are shown in 
Fig~\ref{widths} for a number of CTTS outflows (HL~Tau, HH~30: {\em Ray et 
al.}, 1996; DG~Tau, CW~Tau, and RW~Aur: {\em Dougados et al.}, 2000, {\em 
Woitas et al.}, 2002). Similar results were found by {\em Hartigan et al.}, 
(2004) for the jets from HN~Tau and UZ~Tau~E (see Fig~\ref{widths}). 
These studies show that at the highest spatial resolution currently achieved 
(0\farcs 1 = 14~AU for the Taurus Auriga Star Formation Region), the 
jet is unresolved within 15~AU from the central source. Moreover large opening 
angles (20-30 degrees) are inferred for the  HN~Tau and UZ~Tau~E jets
on scales of 15-50 AU, indicating widths at the jet source 
$<$5 AU ({\em Hartigan et al.}, 2004).  Beyond 50 AU, HH jets seem to slowly 
increase in width with much smaller opening angles of a few degrees. 
FWHM of 20-40 AU are inferred at projected distances of around 100 AU from 
the central source. Such measurements demonstrate that collimation is 
achieved on scales of a few tens of AU, i.e.\ close to the YSO, and 
appear to rule out pure hydrodynamic models for their focusing (see, 
for example, {\em Frank and Mellema}, 1996).  

Here we note that synthetic predictions of jet widths, taking into account
projection and beam dilution effects, have been computed for 
self-similar disk wind solutions (e.g., {\em Ferreira}, 1997 and {\em Casse 
and Ferreira}, 2000). The variation of jet diameter with distance 
from the source is consistent with disk wind models of moderate to high  
efficiency (\.{M}$_{\rm eject}$/\.{M}$_{\rm acc} >$ 0.03) 
(e.g., {\em Garcia et al.}, 2001, {\em Dougados et al.}, 2004 and 
Fig~\ref{widths}). Here efficiency 
is defined as the ratio of the total bipolar jet mass flux to the 
accretion rate. 

\section{KINEMATICS: VELOCITY PROFILES, ROTATION, ACCELERATION AND IMPLICATIONS}

\begin{figure*}
\centering
 \includegraphics[width=16cm]{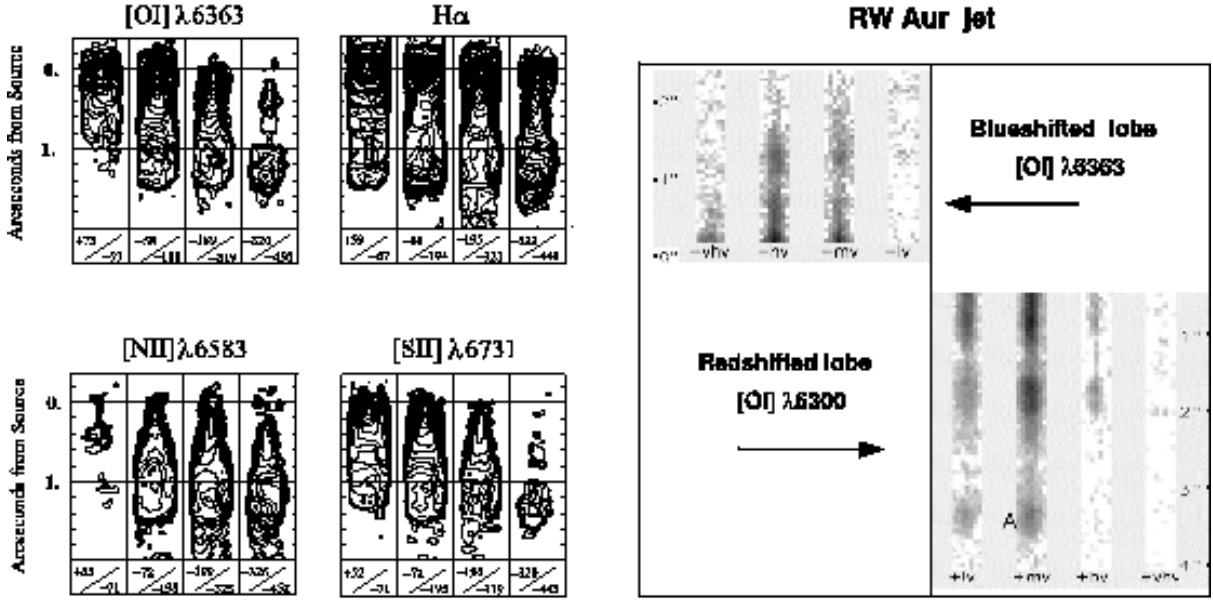}
 \caption{2-D velocity `channel maps' of the blueshifted
jet from DG~Tau (left) and of the  bipolar jet from
RW~Aur (right), reconstructed from HST/STIS multi-slit optical spectra.
For DG~Tau, the low, medium, high and very high velocity intervals are 
approximately from +60~--~-70\kms (LV), -70~--~-195\kms (MV), 
-195~--~-320\kms (HV), and -320~--~-450\kms (VHV) respectively. For RW~Aur, 
velocity bins of about 80 \kms were used starting from -5 \kms 
in the approaching lobe and from +11 \kms in the receding lobe.
Note the increase in jet collimation with increasing velocity.}
\label{chanmaps} 
 \end{figure*}

Spectra of CTTS with jets frequently reveal the presence of two or more 
velocity components ({\em Hartigan et al.}, 1995). In long-slit spectra 
with the slit oriented along 
the jet direction, the so-called high-velocity component (HVC), 
with velocities as large as a few hundred \kms, appears more extended
and of higher excitation than the low-velocity component (LVC), 
which has velocities in the range 10--50 km s$^{-1}$
({\em Hirth et al.}, 1997; {\em Pyo et al.}, 2003).

In order to understand the nature of such components, 
the region of the jet base (first few hundred AU)
has to be observed with sufficient spectral resolution. 
This can be done using integral field spectroscopy, ideally employing AO, 
to produce 3-D data cubes (2-D spatial, 1-D radial velocity).
Slices of the data-cubes then give 2-D images of a jet at different 
velocity intervals, akin to  the `channel maps' of radio interferometry. 
For example {\em Lavalley-Fouquet et al.,} (2000) used OASIS to map 
the kinematics of the DG~Tau jet with 0\farcs 5 resolution 
(corresponding to 70~AU at the distance of 
DG~Tau). Even better spatial resolution, however, can be achieved 
with HST although a long-slit, rather than an integral field spectrograph
has to be employed ({\em Bacciotti et al.}, 2000; {\em Woitas et al.}, 
2002). In particular the jets from RW~Aur and DG~Tau were studied with 
multiple exposures of a 0\farcs 1 slit, stepping the slit transversely 
across the outflow every 0\farcs 07. Combining the exposures together, in 
different velocity bins, again provided channel maps (see Fig. 
\ref{chanmaps}). In such images, the jets show at their base an onion-like 
kinematic structure, being more collimated at higher 
velocities and excitation. The high velocity `spine' 
can be identified with the HVC. 
The images also show, however, progressively wider, slower and less 
excited layers further from the outflow axis, in 
a continuous transition between the HVC and the LVC.
At a distance of about 50-80 AU from the source, though, the low-velocity 
material gradually disappears, while the axial HVC is seen to 
larger distances. The flow as a whole thus {\em appears} to accelerate on 
scales of 50-100 AU.
This structure has recently been shown to extend to the external 
H$_2$-emitting portion of the flow ({\em Takami et al.}, 2004).
All of these  properties  have been predicted by theoretical 
models of  magneto-centrifugal winds (see, for example, the chapters by 
{\em Shang et al.} and {\em Pudritz et al.}).  

The most exciting finding in recent years, however, has been the 
detection of radial velocity asymmetries that could be interpreted as 
{\em rotation} of YSO jets around their axes.
Early hints of rotation were found in the HH~212 jet 
at large distance ($\approx$10$^4$ AU) from the source
by {\em Davis et al.,} (2000). More recently 
indications for rotation have been obtained  
in the first 100 - 200 AU of the jet channel
through high angular resolution observations, 
both from space and the ground ({\em Bacciotti et al.}, 2002; 
{\em Coffey et al.}, 2004; {\em Woitas et al.}, 2005).
\begin{figure*}
\centering
 \includegraphics[width=16cm]{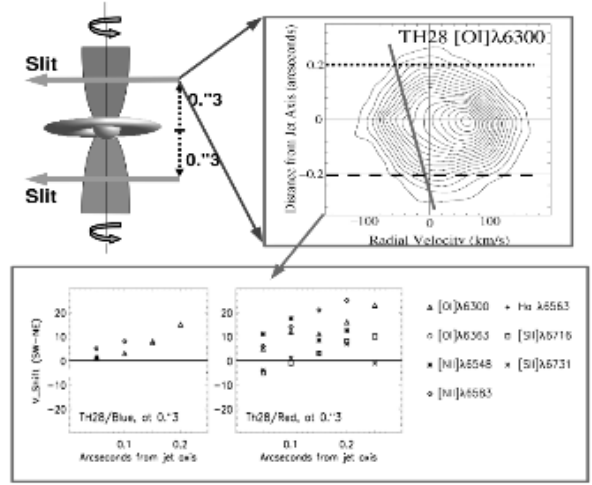}
 \caption{Transverse velocity shifts in the optical emission lines 
detected with HST/STIS across the jets from DG~Tau (Left)  
and Th~28 (Right), at about 50 - 60 AU from the source and 20 - 30 AU 
from the outflow axis. The application of gaussian fitting and 
cross-correlation routines to line profiles from diametrically opposite 
positions centred on the jet axis revealed velocity shifts 
of 5--25 km s$^{-1}$. The values obtained suggest toroidal speeds of
10-20 \kms at the jet boundaries. 
} 
\label{OPTrotation}
 \end{figure*}
\begin{figure*}
\centering
 \includegraphics[width=16cm]{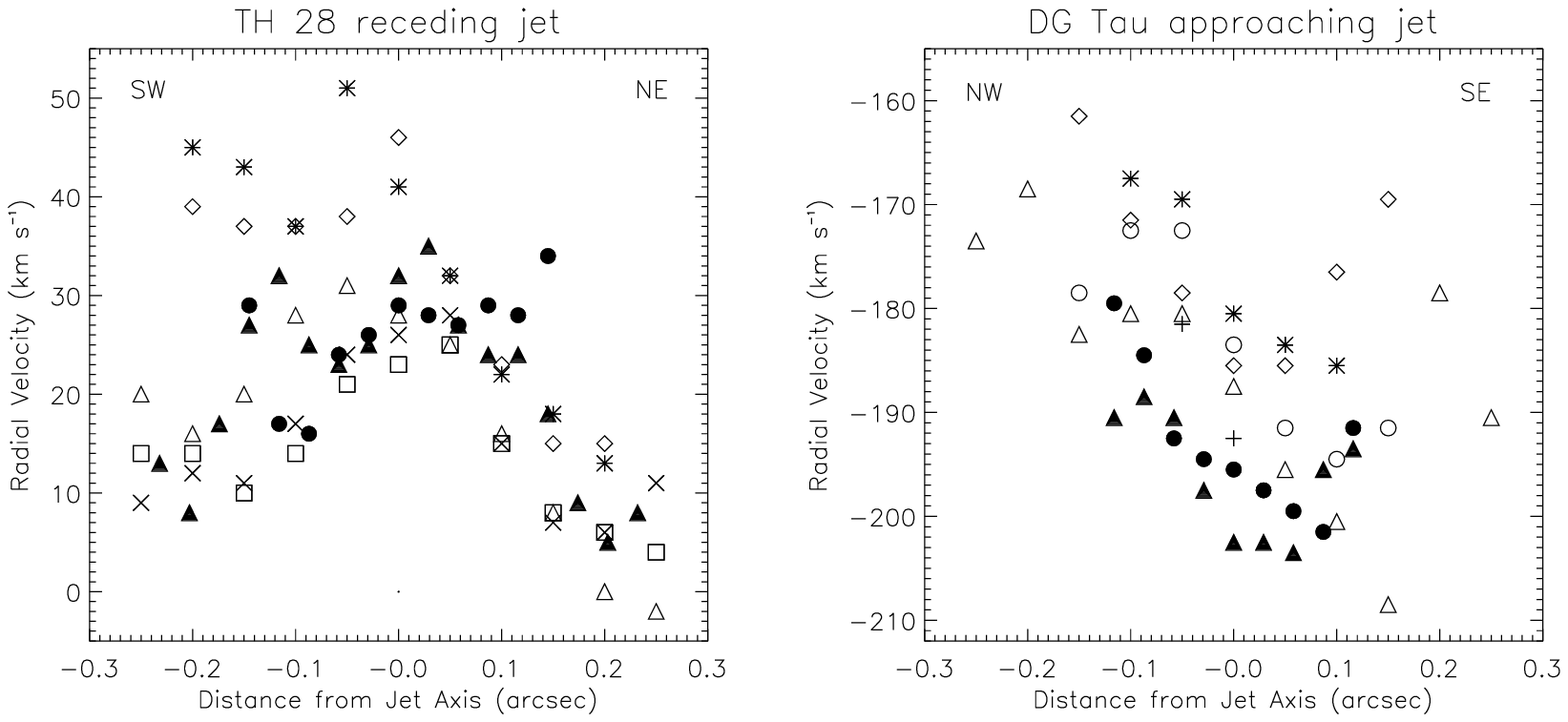}
 \caption{
Radial velocities across the jets from Th~28 (left)
and DG~Tau (right), in the NUV (solid symbols) and optical (hollow
symbols) lines, from HST/STIS spectra taken with the slit transverse to the 
outflow and at 0\farcs 3 from the star. The two datasets fit well together.
The asymmetry in radial velocity from opposing sides of the jet axis, 
seen in both wavelength regimes, suggests rotation.
}
\label{NUVrotation}
 \end{figure*}
Such velocity asymmetries have been seen 
in all the T Tauri jets observed with HST/STIS
(DG~Tau, RW~Aur, CW~Tau, Th~28, and HH~30),
in different emission lines and using slit orientations 
both parallel and perpendicular to the outflow axis. 
For example at optical wavelengths, systematic shifts in radial velocity,
typically from 5 to 25 $\pm$5 \kms , were found at jet positions
displaced symmetrically with respect to the outflow
axis, at 50 - 60 AU from the source 
and 20 - 30 AU from the axis (see Fig. \ref{OPTrotation}). 
Note that the resolving power of STIS in the optical is around 55~\kms . 
Applying, however, gaussian fitting and cross-correlation routines to the 
line profiles, it is possible to detect velocity shifts as small as
 5~\kms .
It should also be mentioned that the sense and degree of 
velocity asymmetry suggesting rotation were found to be consistent in
different elements of various systems (e.g., both lobes 
of the bipolar RW Aur jet, 
the disk and jet lobes in HH~212 and DG Tau) and 
between different datasets ({\em Testi et al.}, 2002; 
{\em Woitas et al.}, 2005; {\em Coffey et al.}, 2005). 

Very recently, such findings have been confirmed by the detection of 
systematic radial velocity shifts 
in the Near Ultra-Violet (NUV) lines of Mg$^+$~$\lambda\lambda$ 
2796, 2803 ({\em Coffey et al.}, in preparation). 
Such lines are believed to arise from the 
fast, highly excited axial  portion of the flow, 
and can be studied with higher angular resolution. 
Unfortunately, however, because of the failure of STIS in August 2003, 
it was only possible to study the jets from Th~28 and DG~Tau in the NUV. 
As expected, the measurements of radial velocity in the NUV lines 
show slightly higher velocities in the axial region
(see Fig. \ref{NUVrotation}). 
Once again asymmetry in radial velocity across the jet is found. 
Assuming this is rotation, both the sense of rotation and its amplitude 
agree with the optical values (see Fig. \ref{NUVrotation}). 

Finally, velocity asymmetries compatible with jet rotation have
also been detected from the ground using ISAAC on the VLT
in two small-scale jets, HH~26 and HH~72, emitting in the  
H$_2$~2.12$\mu$m line ({\em Chrysostomou et al.}, 2005).
The position-velocity diagrams, again based on slit positions transverse
to the outflow, indicate rotation velocities of 5 - 10 \kms at 
$\approx$~2-3$^{\prime\prime}$ from the source and $\approx$1$^{\prime\prime}$
from the jet axis. Note that these jets are driven by Class I sources, 
suggesting rotation is present, as one would expect, from the earliest 
epochs. 

The detection of rotation is interesting {\em per se}, 
as it supports the idea that jets are centrifugally launched
presumably through the action of a magnetic `lever-arm'. 
Assuming that an outflow can be described
by a relatively simple steady magneto-hydrodynamic wind model, the
application of a few basic equations, 
in conjunction with the observed velocity shifts, 
allows us derive a number of interesting 
quantities that are not yet directly observable. 
One of these is the {\em ``foot-point radius''}, 
i.e.\ the location in the accretion disk from where the observed portion of 
the jet is launched ({\em Anderson et al.}, 2003). 
The observations described here are consistent with foot-point radii between 
0.5 and 5~AU from the star, with the NUV (NIR)-emitting layers coming from 
a location in the disk closer (farther) from the axis than 
the optical layers. These findings suggest that at least some of the 
jet derives from an extensive region of the disk (see also {\em Ferreira et
al.}, 2006 who also show intermediate sized magnetic lever arms are favoured). 
It has to be emphasised however that the current observations
only probe the outer streamlines ({\em Pesenti et al.}, 2004) and accurate 
determination of the full transverse rotation profile is critical
to constrain the models. One cannot, for example, exclude the presence 
of an inner X-wind at this stage, as
the spatial resolution of the measurements is not yet sufficient to 
probe the axial region of the flow corresponding to any X-wind ejecta 
(but see {\em Ferreira et al.}, 2006). 
Moreover is not absolutely certain that flow lines can be 
traced back to unique foot-points. In fact, once the region beyond  
the acceleration zone is reached (typically a few AU above the disk),
the wind is likely to undergo various kinds of MHD instabilities that 
complicate the geometry of the field lines. Thus a one-to-one mapping to 
a precise foot-point cannot be expected {\em a priori} (see also below for 
alternative interpretations of the velocity asymmetries).

Assuming however that the current foot-point determinations 
are valid, these can be used to get information on the geometry of the magnetic 
field. In fact from the toroidal and poloidal components of the velocity 
field one can derive the ratio of the corresponding components of the 
magnetic field ${\bf B}$. It is found that at the observed locations  
$B_{\phi}/B_p \sim 3 - 4$ ({\em Woitas et al.}, 2005). 
A prevalence of the toroidal field component at 50 - 100 AU from the star
is indeed predicted by the models that attribute the collimation of the flow 
to a magnetic `hoop stress' ({\em K\"onigl and Pudritz}, 2000). The most 
important quantity derived from the observed putative rotation, 
however, is the amount of angular momentum carried by the jet. 
In the two systems for which we had sufficient information
(namely, DG~Tau and RW~Aur), we have verified that this is
between 60\% and 100\% of the angular momentum that the inner 
disk has to loose to accrete at the observed rate. Thus, the fundamental 
implication of the inferred rotation is that jets are likely to be
the major agent for extracting excess angular momentum from the inner disk and 
in fact this could be their {\em raison d'\^etre.}

It should be noted how a few recent studies have proposed
alternative explanations for the observed velocity asymmetries. 
For example, they could be produced in asymmetric shocks generated 
by jet precession ({\em Cerqueira et al.}, 2006) or by interaction with a 
warped disk ({\em Soker}, 2005).
It seems unlikely, however, that these alternative models could explain
why the phenomenon is so common (virtually all observed jets were found to 
``rotate'') and why the amplitude of the velocity asymmetry is in the
range predicted by disk wind theory. 

Moreover while the above developments have shed new light on 
how jets are generated many puzzles remain. 
For example, a recent study of the disk around RW~Aur, 
suggests the disk rotates {\em in the opposite 
sense} to both of its jets 
({\em Cabrit et al.}, 2006). Although such observations might be explained
by complex interactions with a companion star, it is clear that  
further studies of jets and disks at high spatial/spectral 
resolution are needed to definitely confirm the detection of jet rotation.

Another problem is the observed asymmetry in ejection velocity between 
opposite lobes in a number of outflows. One `classical' example is the 
bipolar jet from RW~Aur, in which the blueshifted material 
moves away from the star with a radial velocity $v_{\rm rad} \sim$ 190 \kms , 
while the redshifted lobe is only moving at $\sim$ 
100-110 \kms . In addition, recent measurements 
({\em L{\'o}pez-Mart{\'i}n et al.}, 2003) have shown that the 
proper motions of the knots in the blue and redshifted lobes are in 
the same ratio as the radial velocities. This suggests that measured 
proper motions represent true bulk motions rather than some form 
of wave. Such findings beg the obvious question: If there are differences 
in parameters like jet velocity (see also {\em Hirth et al.}, 1994 for other 
cases) in bipolar jets, are there differences in even more fundamental 
quantities such as mass and momentum flux? 

What is the origin of the detected near-infrared H$_{\rm 2}$ emission at 
the base of flows? This apparently comes 
from layers just external to the jet seen in forbidden lines.
Several such `H$_{\rm 2}$ small-scale jets' have been found recently,  as, e.g.,
HH 72, HH 26, HH 7-11, but H$_2$ winds are 
also associated with well~known jets from T~Tauri 
stars, as, for example, in DG~Tau. In the latter case 
the emission originates from a warm ($T\sim 2000$ K)
molecular wind with a flow length and
width of 40 and 80~AU, respectively. and has a radial velocity of
$\sim$ 15 km s$^{-1}$ ({\em Takami et al.}, 2004). 
It is not clear if such a molecular component 
is entrained by the axial jet, or if it is a slow external component
of the same disk wind that generates the fast jet. The latter flow geometry 
would again agree with model predictions of magneto-centrifugal driven winds.

\section{LINE DIAGNOSTICS} 

The numerous lines emitted by stellar jets, for example from
transitions of O$^0$, S$^+$, N$^+$, Fe$^+$, H, H$_2$,  
provide a wealth of useful information. By comparing 
various line ratios and intensities with radiative 
models, one is able to determine the basic physical 
characteristics of jets. Not only can we plot the variation 
of critical parameters such as temperature, 
ionization, and density in outflows close to their source and 
in different velocity channels, but also such fundamental quantities as 
elemental abundances and mass flux rates. Recently spectral diagnostic 
methods have been extended from the optical into the near infrared. This 
presents the prospect of not only probing jet conditions in very low 
velocity shock regions but also gives the opportunity of investigating 
more embedded jets from less evolved sources. These studies are 
complementary to the spectral analysis 
aimed at investigating the disk/star interaction zone and the role of 
accretion in determining outflow properties.

Determination of the line excitation mechanism in T Tauri stars
is a long-standing issue. It is critical in particular for a
detailed comparison of wind model predictions with observations.
Comparison of the optical line emission properties of the DG~Tau and
RW~Aur jets with what is expected from three different classes of
excitation mechanism (mixing layers, ambipolar diffusion, planar
shocks) show that line ratios in these jets are best
explained by shock excitation with moderate to large velocities (50-100 \kms)
({\em Lavalley-Fouquet et al.}, 2000; {\em Dougados et al.}, 2003) indicating
that time variability plays a dominant role in the heating process.

Since Protostars and Planets~IV, line diagnostics of stellar jets have 
developed to the point that not only can we determine the usual quantities
such as the electron density ($n_e$) and temperature ($T_e$), 
but a number of additional ones as well. The most important of which 
is probably {\em total density}, $n_H$. In fact
all models for the dynamics and radiative properties of jets 
are highly dependent on this parameter, either directly or through 
derived quantities such as the jet mass or angular momentum fluxes 
(see Section 3 and Section 5). 
\begin{figure}[ht]
\centering
 \includegraphics[width=7cm]{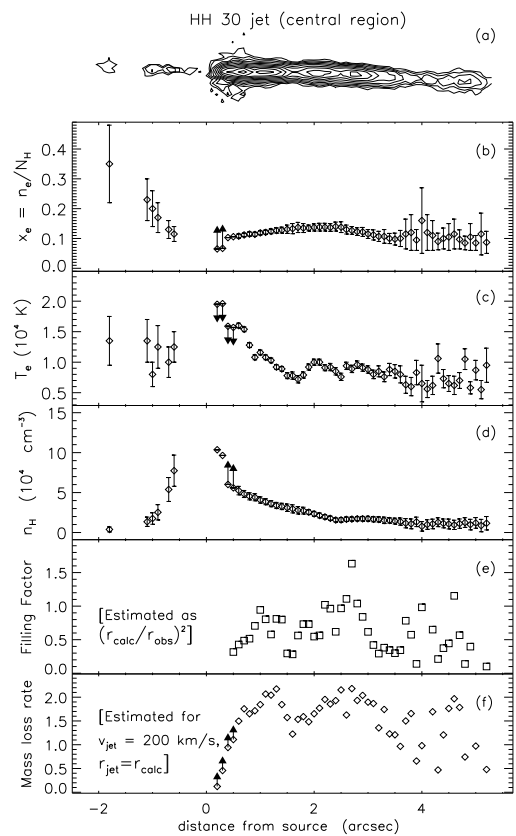}
 \caption{  
Physical quantities along the HH30 jet derived using the BE technique
(see text) from HST/WFPC2 narrow-band images.
From top to bottom: [SII] emission, ionization fraction, electron 
temperature, total density, filling factor and mass flux
in units of 10$^{-9}$ M$_{\odot}$ yr$^{-1}$ along the jet.} 
\label{hh30_jet}
 \end{figure}
\begin{figure}[ht]
\centering
 \includegraphics[width=8cm]{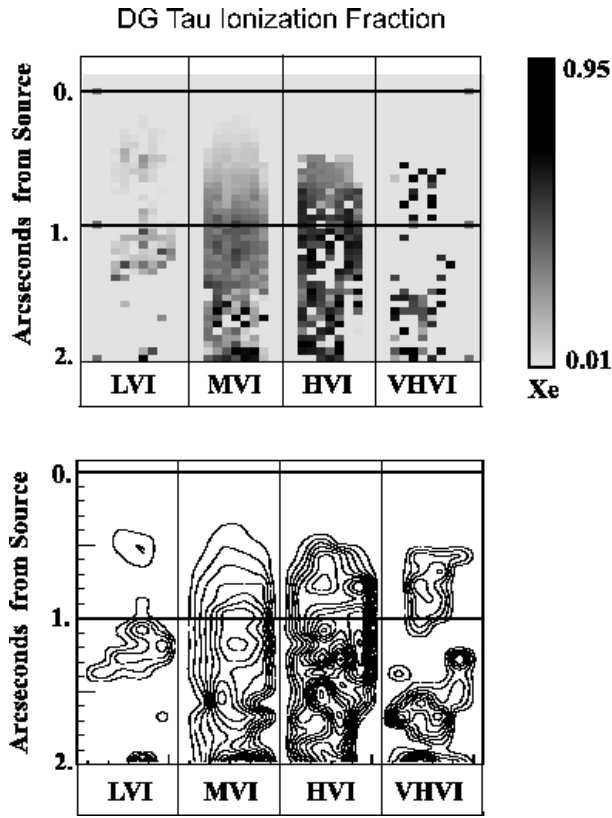}
 \caption{2D maps of the level of ionization in the first 200 AU 
of the DG~Tau jet in the different velocity channels indicated
in Fig.\ \ref{chanmaps}. Ionization values
were derived by applying the BE technique to HST/STIS multiple spectra.
} 
\label{physmaps}
 \end{figure}

In early studies, as now, physical quantities, such as 
$n_e$ and $T_e$, were determined from line ratios sensitive to 
these parameters. The observations however were made at low spatial 
resolution and thus effectively integrated over the shock cooling zone
(e.g.\ {\em B\"ohm et al.}, 1980; {\em Brugel et al.}, 1981). 
Estimates of other important quantities, such as the ionization fraction
and the strength of the pre-shock magnetic field, had to wait for 
studies in which the observed line intensities and ratios were 
compared with predicted values 
based on shock models (e.g., {\em Hartigan et al.}, 1994). 
More recently, however, a simpler method, referred to as the `BE' technique 
({\em Bacciotti and Eisl\"offel}, 1999) has been developed to measure 
$x_e = n_e/n_H$ and $T_e$ directly from readily observed optical 
line ratios utilising transitions of O$^0$, N$^+$ and S$^+$
({\em Bacciotti and Eisl\"offel}, 1999; {\em Bacciotti}, 2002).  
The procedure is based on the fact that the gas producing forbidden lines is 
collisionally excited and no assumptions are made as to the heating 
mechanism. Such an approach is clearly expedient although not a substitute
for a detailed model. The method assumes for example that  
the emitting gas is at a single temperature, which is not true in the 
cooling zone behind a shock. That said, one should consider
parameters derived from the BE technique as relevant to those regions 
behind a shock in which the used lines peak in their emission   
(see the discussion and diagrams in {\em Bacciotti and Eisl\"offel}, 1999).

The basic premise of the BE method is that Sulphur is singly ionised in YSO 
jets because of its low ionization potential. At the same time the 
ionization fraction of Oxygen and Nitrogen is assumed to 
be regulated by charge-exchange with Hydrogen (and recombination). 
This provides the link between observed line ratios and $x_e$. The 
necessary conditions for these assumptions are generally fulfilled in
YSO jets. Since however photo-ionization is ignored, applicability 
of the BE technique is limited to regions far away from any strong  
sources of UV photons, such at the vicinity of terminal bow-shocks. 
One must also assume a set of elemental abundances, 
appropriate for the region under study (see the discussion in 
{\em Podio et al.}, in preparation).

A number of jets have been analysed with this technique using moderate 
resolution long-slit spectra and integral field spectroscopy 
({\em Bacciotti and Eisl\"offel}, 1999; {\em Lavalley-Fouquet et al.}, 2000;
{\em Dougados et al.}, 2000; {\em Nisini et al.}, 2005; {\em Podio et al.}, 
in preparation; {\em Medves et al.}, in preparation).  Typical 
$n_e$ values are found to vary between 
50~cm$^{-3}$ and 3.10$^{3}$~cm$^{-3}$, $x_e$ to range from 0.03 to 0.6, 
and $T_e$ to decrease along a jet from a peak of 2--3 10$^{4}$~K close to 
the source to averages of 1.0--1.4 10$^{4}$~K on larger scales. 
The variation of $n_e$ and $x_e$ along the flow {\em is consistent with 
ionization freezing close to the source}, followed by 
slow non-equilibrium recombination. The mechanism that produces 
a high degree of ionization close to, although not coincident with,
the base of the jet (see Fig. \ref{physmaps}), however, is not known. 
It might, for example, derive from a series of spatially compact 
non-stationary shocks at the jet base ({\em Lavalley-Fouquet et al.}, 2000; 
{\em Massaglia et al.}, 2005; {\em Massaglia et al.}, in preparation). 
Whatever the mechanism a comparison of 
various jets seems to suggest that it has similar efficiencies in all cases: 
lower ionization fractions are 
found in the densest jets, i.e.\ those that recombine faster. In any 
event, the realization that stellar jets are only 
partially ionised  has provided new, more accurate estimates of the 
total density $n_H$. These estimates are much bigger than previous ones 
and typically, on large scales, range from 10$^3$ to 10$^5$~cm$^{-3}$.  
This, in turn, implies that jets can strongly affect their environments, 
given their markedly increased mass, energy and momentum fluxes. 
Taking into account typical emissivity filling factors, mass loss 
rates are found to be about 10$^{-8}$ - 10$^{-7}$  M$_\odot$ yr$^{-1}$
({\em Podio et al.}, in preparation) for CTTS jets.
The associated linear momentum fluxes (calculated as $\dot{P} = 
v_{jet} \dot{M}$) are higher, or of the same order, as those measured in 
associated coaxial molecular flows, where present. This suggests that 
partially ionised YSO jets could drive the latter.  

The BE method is well suited to analysing large datasets, 
such as those provided by high angular resolution observations. An example of 
its application, to HST narrow-band images of the HH~30 jet, 
is shown in Fig.\ \ref{hh30_jet} ({\em Bacciotti et al.}, 
1999). A similar analysis of spectra taken from the ground using AO
is presented in {\em Lavalley-Fouquet et al.}, (2000).
Even more interesting, is the application of the technique to 
the 2-D channel maps reconstructed from parallel slit STIS 
data (see Fig. \ref{chanmaps}). In this way one can obtain
high angular resolution 2-D maps of 
the various quantities of interest in the different velocity channels, 
as illustrated in Fig. \ref{physmaps} for the ionization fraction in the 
DG~Tau jet ({\em Bacciotti}, 2002; {\em Bacciotti et al.}, in preparation).
The electron density maps, for example, 
confirm that $n_e$ is highest closest to the star,
nearer the axis, and at the highest velocities. 
At the jet base, one typically finds 0.01$< x_e<$ 0.4, 
and total densities up to  10$^6$ cm$^{-3}$. 
In the same region 8.10$^3$ $< T_e<$ 2~10$^4$~K. 
These values can be compared with those predicted by MHD
jet launching models. Finally, from $n_H$, the jet diameter
and the de-projected velocity, one can determine 
the initial mass flux in the jet $\dot{M}_{\rm jet}$. Typical 
values are found to be around 10$^{-7}$~M$_{\odot}$~yr$^{-1}$, with the 
colder and slower external layers of the jet contributing most to the 
flux. Such values can be combined with known accretion rates in these
stars to produce $\dot{M}_{\rm jet}$/$\dot{M}_{\rm acc}$ ratios. Note that 
accretion rates are determined independently through line veiling ({\em 
Hartigan et al.}, 1995). Typical ratios in the range 0.05 -- 0.1 are found. 
This seems inconsistent with cold disk wind models although warm disk winds 
with moderate magnetic lever arms are expected to produce such ratios 
({\em Casse and Ferreira}, 2000).

In the last few years application of line diagnostics has been 
extended from the optical into the near infrared
({\em Nisini et al.}, 2002; {\em Pesenti et al.}, 2003; {\em Giannini et al.}
2004; {\em Hartigan et al.}, 2004). As stated at the beginning 
of this section, this presents the 
prospect of not only probing jet conditions in very low velocity shock 
regions but also the possibility of investigating more embedded, and 
less evolved jets. For example, using NIR lines of Fe$^+$, 
one can determine not only the electron density in denser embedded 
regions of the jet, but also such fundamental quantities as the 
visual extinction A$_{\rm V}$ along the line of sight to an outflow. 
Note that A$_{\rm V}$ has to be known if we are to correct line ratios 
using lines that are far apart in wavelength.
In addition, the NIR H$_2$ lines provide a probe of excitation conditions 
in low velocity shocks near the base of the flow where some molecular 
species survive.

Very recently, the potential of a combined optical/NIR set of line diagnostics
has been exploited ({\em Nisini et al.}, 2005; {\em Podio et al.}, in 
preparation), using a variety of transitions in the 0.6 to 2.2\ts$\mu$m range. 
This approach has turned out to be a very useful means of determining 
how physical quantities vary in the different stratified layers 
behind a shock front as well as providing additional checks on many 
parameters. For example, the combination of red and NIR Fe$^+$ lines 
gives an independent estimate of  
T$_e$ and n$_e$, which does not rely on the choice of elemental 
abundances. The electron density and temperature, derived from Iron lines,  
turn out to be higher and lower respectively, than those determined 
from optical lines. This is in agreement with the prediction that Iron 
emission should, on average, come from a region within the post-shock cooling 
zone that is farther from the shock front than those regions giving rise 
to the optical lines. Hence it is cooler and of higher 
density. Moreover analysis of NIR Ca$^{\rm +}$ and C$^{\rm 0}$ 
lines show that jets possess regions of even higher density 
(n$_{\rm H}$ up to 10$^6$~cm$^{-3}$).
Finally, various lines have been used to estimate the depletion onto 
dust grains of Calcium and Iron with respect to solar values. The amount
of depletion turns out to be quite substantial: around 30-70\% for Ca and 
90\% for Iron. This leads to the suggestion that the weak shocks 
present in many jets are not capable of completely destroying 
ambient dust grains, as expected theoretically ({\em Draine}, 1995). 


\section{PROBING THE OUTFLOW ON AU SCALES: USE OF SPECTRO-ASTROMETRY}

As this review (and others in this volume) shows,
there has been a dramatic improvement over the past decade in the spatial
information provided by modern instrumentation, with arguably the most
detail delivered by the HST (e.g., {\em Bacciotti et
al.}, 2000). AO also played an important role in allowing
us to peek closer to the YSO itself (e.g., {\em Dougados et al.}, 2002).
Nevertheless, it still remains true that the spatial resolution of
astronomical observations are constrained by the diffraction limit of the
telescope (as well as optical aberrations and atmospheric seeing in many 
cases). 
In order to probe down to the central engine, we need to achieve a resolution
of $\sim$\,10\,AU or better which corresponds to $\lesssim\,0.05\arcsec$
for the nearest star forming regions. In other words, we need milliarcsecond 
resolution!

Here we briefly describe the technique of spectro-astrometry and how it
can be used to determine the spatio-kinematic structure of sources well
below the diffraction limit of the telescope, and then review some of the
important results obtained using this technique.

The great value and appeal of spectro-astrometry lies in its rather
simple application and the fact that it does not require specialist
equipment nor the best weather conditions. With no more than a standard
CCD in a long-slit spectrograph, we have in return a tool which is capable of 
probing emission structures within AU scales of the YSO. An unresolved star 
is observed with a long-slit spectrograph and the positional centroid
of the emission along the slit is determined as a function of wavelength.
If the unresolved source consists of an outflow or binary with distinctive
features in their spectra, the centroid position will shift relative to the
continuum at the wavelength of the features. As such, the only factor
which affects the accuracy of this technique is the ability to measure an
accurate centroid position and this is ultimately governed by the number
of photons detected and a well sampled seeing profile (hence a small pixel
scale, with the added requirement of a uniform CCD). Each photon detected
should be within a distance, defined by the seeing disk, from the centroid
and the uncertainty in this position reduces by $\sqrt{N}$ when $N$
photons are detected. This results in a position spectrum whose accuracy,
measured in milliarcseconds, is given by $\Delta x
\sim 0.5 \times \textrm{FWHM}_\textrm{seeing} / \sqrt{N}$ ({\em Takami et al.}, 
2003). Consequently, 
brighter emission lines are better suited to probing the inner regions 
of the YSO.

As {\em Bailey}, (1998) explains, the method is not new and has been used in
previous works, although these involved specialist techniques and
equipment ({\em Beckers}, 1982; {\em Christy, Wellnitz and Currie}, 1983). 
Moreover a number of authors (e.g., {\em Solf and B\"ohm}, 
1993 and {\em Hirth et al.}, 1997) have used long-slit spectroscopy to examine
sub-arcsecond kinematic structure of the HVC and LVC in a number of outflows. 
{\em Bailey}, (1998) revived interest in the technique by
demonstrating that one could routinely recover information on milliarcsecond
scales by taking long-slit spectra at orthogonal and anti-parallel position
angles, i.e.\ by rotating the instrument and pairing spectra taken at 
$0^\circ\ \&\ 180^\circ$ and $90^\circ\ \&\ 270^\circ$ apart. Each
anti-parallel pair is subtracted to remove any systematic errors introduced 
by the telescope and/or instrument, as such signals should be independent of 
rotation whereas the true astronomical signal is reversed (see also 
{\em Brannigan et al.}, in preparation).

\begin{figure*}[ht]
\includegraphics[width=17cm]{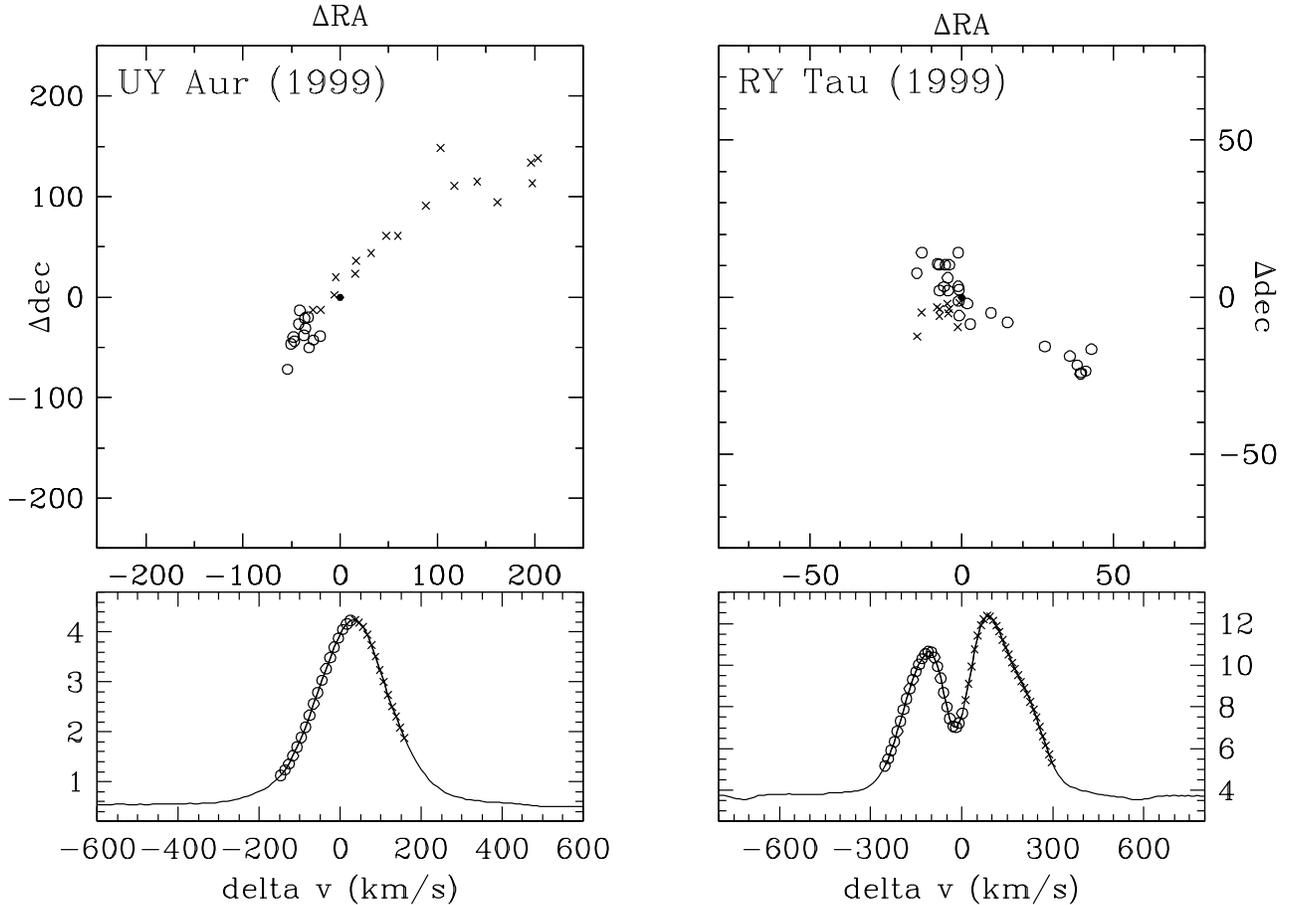}

\caption{Example spectro-astrometry data of the H$\alpha $ line. The 
upper panels show x-y plots of the spectro-astrometric offsets at each 
position across the line profile (circles and crosses for blue- and red-shifted 
components, respectively). The integrated line profile is shown below. The data 
were taken using the ISIS instrument on the William Herschel Telescope 
on La Palma. \textit{Left}: Detection of a 
\textit{bipolar} outflow in UY~Aur implying the presence of a disk gap in this 
object. 
\textit{Right}: The same for RY~Tau although the 
evidence suggests either a binary companion or a mono-polar jet 
with the red-shifted component hidden from view.}
\end{figure*}

\subsection{Spectro-astrometry -- the science}
The first extensive use of spectro-astrometry came with the study of 
{\em Bailey}, (1998) who used it to prove the technique on known binaries, in 
the course of which two previously unknown binaries were also discovered. 
Since then, {\em Garcia et al.}, (1999) and {\em Takami et al.}, (2001, 2003) 
have surveyed a number of YSOs revealing jets on 
scales of $\sim 10$\ AU from the source in a few cases. Perhaps most 
interesting, are discoveries of \textit{bipolar} jets at these small scales. 

Forbidden lines only trace outflows in those zones with less than the critical 
electron density for the line. As jets tend to have higher densities
closer to their source, this means that a point is reached where individual 
forbidden lines 
fade. In contrast, a permitted line traces activity all the way to the star. 
Such lines, e.g.\ H$\alpha$, however not only map outflows but magneto-spheric 
accretion onto the YSO as well (e.g., {\em Muzerolle et al.}, 1998). It follows 
that the profile of a permitted line at the star tends to be a 
mixture of outflow and inflow components. As we will show, spectro-astrometry
is a way of unravelling these respective contributions. Now as mentioned
previously, if a spectrum is taken of a CTTS, it tends only to show 
blueshifted forbidden lines. This is readily understandable 
if the forbidden line emission comes only from an outflow and the redshifted 
(counter-flow) at the star is obscured by an accretion disk (a view that 
is also endorsed by spectro-astrometry). Now if we search for a 
spectro-astrometric offset in the blue-shifted wing of a permitted line
we find the maximum offset at the blueshifted jet velocity and 
in the same direction. The offsets however are much smaller than those 
measured for the forbidden lines, implying the permitted emission tends to 
come from much closer to the source. Remarkably {\em Takami et al.,} (2001, 
2003) found bipolar H$\alpha $ emission centred on RU~Lupi and CS~Cha, 
clearly indicating a direct line of sight to the counter-flow through the 
accretion disk in these stars. This is interpreted as evidence of a 
sizable gap or dust hole in the disk at a radius of $\sim 1-5\ \mathrm{AU}$ 
from the protostar. The gap itself could be generated by a 
planetary body (see, for example, {\em Varni\'ere et al.}, 2005) although
it could equally be due to the development of very large dust grains in the 
innermost region of the disk (see, for example, {\em Watson and Stapelfeldt}, 
2004). Such large dust grains would have reduced opacity. 
Supporting evidence for the presence of gaps is seen in the spectral
energy distributions of these objects; they show mid-infrared emission dips 
consistent with temperatures of $\sim\ 
200~\mathrm{K}$, coincident with the ice condensation temperature where the 
increase in surface density may aid planet formation ({\em Boss}, 1995).

Recently the technique has been used in the near-infrared. As well as allowing 
us to investigate younger and more embedded sources, this wavelength range 
also makes available other lines as probes. For example, 
Pa$\beta$~(1.2822\ $\mu$m) is found in the spectra of many T-Tauri stars 
and was believed to exclusively trace accretion. 
{\em Whelan et al.,} (2004) showed however that this is not always 
the case using spectro-astrometry. In particular large spatial offsets 
relative to the YSO were found in the line wings (something which would not 
be expected in the case of accretion). Magneto-spheric accretion models have 
always struggled to explain the detailed profiles of permitted lines 
({\em Folha and Emerson}, 2001). 
Spectro-astrometry seems to have resolved this problem by identifying  
those parts of the line attributable to an outflowing jet. It is also worth
noting that both {\em Takami et al.}, (2001) and {\em Whelan et al.}, 
(2004) show evidence which suggests that offsets from the 
star increases with velocity, consistent with the presence of an acceleration
zone.

Finally, {\em Whelan et al.}, (2005) used spectro-astrometry to report the 
first detection of an outflow from a brown dwarf. Their data suggest many 
similarities (allowing for scaling factors) between the brown dwarf outflow 
and those seen in CTTS. Observations, such as these, suggest a 
universal correlation between the gravitational collapse of an object with 
an accretion disk and the generation of an outflow.

\section{FROM BROWN DWARFS to HERBIG Ae/Be STARS}

It is conceivable that the accretion/ejection mechanism responsible for the
generation and collimation of jets becomes substantially modified, or may 
not even operate, as one goes to sources of substantially higher or 
lower mass than T\,Tauri stars. Thus, it is an interesting question if 
such objects also produce jets and if they are similar to those from 
CTTS. Certainly exploring how outflows vary with 
the mass of the central object (varying escape velocity, radiation field, etc)
can provide useful constraints and tests for any proposed accretion/ejection 
model.

Recent studies have detected signatures of accretion in a wide range of 
objects from brown dwarfs, with masses as low as 0.03\,M$_{\odot}$
({\em Jayawardhana et al.}, 2003; {\em Natta et al.}, 2004; 
{\em Mohanty et al.}, 2005), through Very Low Mass (VLM) stars 
({\em Scholz and Eisl\"offel}, 2004) to Herbig Ae/Be stars 
({\em Finkenzeller}, 1985; {\em B\"ohm and Catala}, 1994; {\em Corcoran 
and Ray}, 1997) of 
2-10~\msun . While the accretion rates show a large spread at any given mass, 
it seems that a relationship \.{M}$_{acc} \propto$ M$^{2}_{obj}$ 
holds for the upper envelope of the distribution (see, for example,
{\em Natta et al.}, 2004). This dependence is much
steeper than expected from viscous disk models with a constant
viscosity parameter $\alpha$, which would predict a much shallower 
relationship ({\em Natta et al.}, 2004; 
{\em Mohanty et al.}, 2005). Strongly varying disk ionisation with the central
object mass caused by its X-ray emission ({\em Muzerolle et al.}, 2003) or disk
accretion controlled by Bondi-Hoyle accretion from the large gas reservoir of
the surrounding cloud core ({\em Padoan et al.}, 2005) have been proposed to
understand the steep relationship between accretion rate and central
object mass.

In order to estimate the minimum mass of a source that we might be able to 
detect an outflow from, it is plausible to extrapolate the known linear  
correlation between ejection and accretion rates. Moreover
{\em Masciadri and Raga}, (2004) have shown that the luminosity of a 
putative brown dwarf jet should scale approximately with its mass
outflow rate. Thus, giving the known accretion rates of brown dwarfs, 
we expect their outflows to be about 100 times fainter at best than those from 
CTTS, i.e.\ just within the 
reach of the biggest telescopes ({\em Whelan et al.}, 2005). 

The first very low mass object, LS-RCrA-1 found to show forbidden
lines typical of outflows from T\,Tauri stars, was discovered by  
{\em Fern\'andez and Comer\'on}, (2001). Its mass has been estimated as 
substellar ({\em Barrado y Navascu\'es et al.}, 2004), but 
the forbidden line emission could not be resolved spatially ({\em 
Fern\'andez and Comer\'on}, 2005). In the same work, however, a
4$\arcsec$-long jet and a 2$\arcsec$-long counterjet were reported on  
from the VLM star Par-Lup3-4. Simultaneously {\em Whelan et al.}, 
(2005) detected a spatially resolved outflow from 
the 60 Jupiter-mass brown dwarf $\rho$\,Oph\,102, using spectro-astrometry. 
The outflow is blue-shifted, by about -50\,km\,s$^{-1}$ with respect to its 
source, and has characteristics similar to those from CTTS. 

Moving to the other end of the mass spectrum, it is also interesting to
investigate if the intermediate mass Herbig Ae/Be stars show accretion/ejection
structures like those of the lower mass T\,Tauri stars.
Several large-scale Herbig-Haro flows have been known for some time, which 
may be driven by Herbig Ae/Be stars. In many cases however it is not 
certain that the Herbig Ae/Be star, as opposed to a nearby, unrelated 
T\,Tauri star or a lower mass companion, could be responsible. Examples include
HH39 associated with R\,Mon ({\em Herbig}, 1968), HH218 associated with 
V645\,Cyg ({\em Goodrich}, 1986), and HH215 and HH315 associated with 
PV\,Cep ({\em Neckel et al.}, 1987; {\em Gomez et al.}, 1997; {\em Reipurth et
al.}, 1997). The first jet that could unambiguously be traced back to a Herbig 
Ae/Be star was HH398 emanating from LkH$\alpha$\,233 ({\em Corcoran and Ray}, 
1998). All these outflow sources exhibit blue-shifted forbidden emission lines
in their optical spectra with typical velocities of a few hundred km\,s$^{-1}$,
similar to CTTS.

More recently, small-scale jets from two nearby Herbig Ae stars have been found 
by coronographic imaging in the ultraviolet with the HST. 
These jets and counterjets from HD163296 (HH409, {\em Devine et al.}, 2000; 
{\em Grady et al.}, 2000) and HD104237 (HH669, {\em Grady et al.}, 2004) 
are only a few arcseconds long. They are readily seen in Ly$\alpha$ as 
the contrast between the star and the shocked flow is
much more favourable than in the optical: Ly$\alpha$/H$\alpha$ $\ge$ 10 
according to shock models (e.g.\ {\em Hartigan et al.}, 1987).

Summarising, we find that Herbig-Haro jets both in the form of small-scale 
jets and parsec-scale flows are ubiquitous in CTTS. They 
seem to
be much rarer towards the higher mass Herbig AeBe stars, and only one clear 
example of an outflow from a brown dwarf is known so far. Nevertheless 
this seems to
indicate that the accretion/ejection structures in all objects are similar 
and the same physical processes are at work over the whole mass spectrum.
The conditions for the generation of jets seem to be optimal in CTTS T\,Tauri 
stars, while under the more extreme environments, both in the lower mass
brown dwarfs and in the higher mass Herbig Ae/Be stars, jet production 
could be less efficient. Current models of the magneto-centrifugal launching
of jets from young stars will have to be tested for the
conditions found in brown dwarfs and Herbig AeBe stars, in order to see if
they are can reproduce the frequency and physical characteristics of the
observed flows.

\section{THE FUTURE: TOWARDS RESOLVING THE CENTRAL AU}

While it is clear from this review that high angular resolution observations
(on scales $\approx$ 0\farcs 1) have provided important information  
on the launch mechanism, the true `core' of the engine lies below the 
so-called `Alfv\`en surface'. This surface is located within a few AU of the 
disk (i.e.\ tens of milliarcseconds for the nearest star forming region) 
so the core cannot be resolved with conventional instrumentation either 
from the ground or space. This, however, is about to change with 
the new generation of optical/NIR interferometers coming on-stream, opening up 
the exciting possibility of exploring this region for the first time. 
Two facilities are being rolled out at present in which Europe will play an 
active role: the VLT Interferometer (VLTI) at ESO-Paranal, and 
the Large Binocular Telescope (LBT) at Mount Graham in Arizona.

The VLTI will operate by connecting, in various combinations, 
up to four 8-m and four smaller (1.8-m) auxiliary telescopes. 
The latter can move on tracks to a number of fixed stations so as to 
obtain good {\em (u,v)} plane coverage. The beams, of course, from the 
various telescopes have to be combined in a correlator. In this regard 
the AMBER instrument is of particular interest to the study of YSO 
jet sources as it allows for medium resolution spectroscopy in the NIR
(e.g.\ high velocity Paschen $\beta$ emission).  AMBER, can combine up 
three AO corrected beams, thus allowing ``closure phase'' to be achieved
and it will provide angular resolution as small as a few milliarcseconds. 
In the early days of operating this instrument, incomplete {\em (u,v)} plane 
coverage is expected and in this case models of the expected emission are 
needed to interpret the observations ({\em Bacciotti et al.}, 2003). The LBT 
Interferometer, in contrast to the VLTI, is formed by combining the beams 
of two fixed 8.4-m telescopes and hence has a fixed baseline. 
Effectively it will have the diffraction limited resolution of a 23-m 
telescope ({\em Herbst}, 2003) and excellent {\em (u,v)} plane coverage because
of the shortness of the baseline in comparison to the telescope apertures.
Moreover it will complement the VLTI through its short projected 
baseline spacings, spacings that are inaccessible to the former. 
The LBT Interferometer will initially operate in the NIR in so-called 
Fizeau mode providing high resolution images over a relatively 
large field of view (unlike the VLTI). A future extension into the optical 
is however planned. 
 
Radio provides a means of 
obtaining high spatial resolution observations of YSO jets close to 
their source if the source is highly embedded ({\em Girart et al.}, 2002). 
Their free-free emission, 
however, tends to be rather weak and so the number of outflows mapped so 
far at radio wavelengths has been quite small. This will change dramatically
when the new generation of radio interferometers, in particular e-MERLIN 
(extended MERLIN) and EVLA (extended VLA), come on-line. Although there will 
be modest improvements in resolution, the primary gain will be in sensitivity 
allowing, in some
cases, the detection of emission 20-50 weaker than current thresholds. This 
increase in sensitivity will be achieved through correlators and telescope 
links with much broader band capacities than before. 
Such improvements will not only lead to the detection of more jet sources, and 
hopefully allow meaningful statistical studies, but perhaps more importantly to 
the detection of non-thermal components in known outflows as already hinted at 
in a number of studies (see {\em Girart et al.}, 2002; {\em Ray et al.}, 
1997; {\em Reid et al.}, 1995). Polarization studies of such emission 
in turn would give us a 
measure of ambient magnetic field strength and direction, parameters that are 
poorly known at present. Finally it is worth noting that a number of studies 
have shown that H$_{\rm 2}$O masers may in some instances be tracing outflows 
({\em Claussen et al.}, 1998; {\em Torrelles et al.}, 2005) and not 
circumstellar disks as often assumed. 
High resolution polarization studies of such masers 
can not only provide information on magnetic field strengths and direction but, 
when combined with multi-epoch studies, information on how the magnetic 
fields evolve with time ({\em Baudry and Diamond}, 1998). 

The other large interferometric facility being planned for observations 
at the sub-mm wavelength range is ALMA, an array of 64 12m antennas
to be built in the Atacama desert in Chile, not far from the  VLTI
site. With ALMA we will be able not only 
to routinely measure disk rotation but also conceiveably rotation in 
molecular outflows if present. The future therefore for this field is very 
bright indeed. 

\bigskip

\textbf{Acknowledgments.} TR, CD, FB and JE wish to acknowledge support
through the Marie Curie Research Training Network JETSET (Jet Simulations, 
Experiments and Theory) under contract MRTN-CT-2004-005592. TR would
also like to acknowledge assistance from Science Foundation Ireland 
under contract 04/BRG/P02741. Finally we wish to thank the referee for 
his very helpful comments while preparing this manuscript.  

\bigskip

\centerline\textbf{ REFERENCES}
\bigskip
\parskip=0pt
{\small
\baselineskip=11pt

\refs Anderson, J. M., Li, Z.-Y., Krasnopolsky, R., and Blandford, R. (2003)
{\em Astrophys. J., 590}, L107-L110.

\refs Bacciotti, F. (2002) {\em Rev. Mex. Astron. Astrofis., 13}, 8-15.

\refs Bacciotti, F. and Eisl\"offel, J. (1999)
{\em Astron. Astrophys., 342}, 717-735.

\refs Bacciotti, F., Eisl\"offel, J., and Ray, T.P. (1999) 
{\em Astron. Astrophys., 350}, 917-927.

\refs Bacciotti, F., Mundt, R., Ray, T.P., Eisl\"{o}ffel, J., Solf, J., and
Camenzind, M. (2000) {\em Astrophys. J., 537}, L49-L52.
 
\refs Bacciotti, F., Ray, T.P., Mundt, R., Eisl\"offel, J., and 
Solf, J. (2002) {\em Astrophys. J., 576}, 222-231.

\refs Bacciotti, F., Testi, L., Marconi, A., Garcia, P.~J.~V., Ray, T.~P., 
Eisl{\"o}ffel, J., and Dougados, C. (2003) {\em Astrophys. Sp. Sci., 286}, 
157-162.

\refs Bailey, J.A. (1998) {\em Mon. Not. Roy. Astr. Soc., 301}, 161-167.

\refs Barrado y Navascu\'es, D.,  Mohanty, S., and Jayawardhana, R. (2004) 
{\em Astroph. J., 604}, 284-296.

\refs Baudry, A.\ and Diamond, P.~J.\ (1998) 
{\em Astron. Astrophys., 331}, 697-708. 

\refs Beckers, J. (1982) {\em Opt. Acta, 29}, 361-362.

\refs B\"ohm, T. and Catala, C. (1994) {\em Astron. Astrophys., 290}, 167-175.

\refs B\"ohm, K. H., Mannery, E., and Brugel, E. W. (1980)
{\em Astrophys. J., 235}, L137-L141. 

\refs Boss, A. (1995) {\em Science, 267}, 360-362.

\refs Brugel, E. W., B\"ohm, K. H., and Mannery, E. (1981) 
{\em Astrophys. J. Suppl., 47}, 117-138. 

\refs Cabrit, S., Edwards, S., Strom, S.~E., and Strom, K.~M.\ 1990, 
{\em Astrophys. J., 354}, 687-700 

\refs Cabrit, S., Pety, J., Pesenti, N., and Dougados, C. 
(2006)  {\em Astron. Astrophys.}, in press.

\refs Casse, F., and Ferreira, J.\ 2000, 
{\em Astron. Astrophys., 353}, 1115-1128

\refs Cerqueira, A. H., Velazquez, P. F., Raga, A. C., Vasconcelos, M. J., 
and De Colle, F. (2006) {\em Astron. Astrophys.}, in press.

\refs Christy, J., Wellnitz, D., and Currie, D. (1983) In {\em Current 
Techniques in Double and Multiple Star Research} R.\ Harrington and O.\ 
Franz,  eds.) (IAU Coll. 62), {\em Lowell Obs. Bull., 167}, 28-35.

\refs Chrysostomou, A., Bacciotti, F., Nisini, B., Ray, T.~P., 
Eisl{\"o}ffel, J., Davis, C.~J., and Takami, M.\ (2005) 
In {\em PPV Poster Proceedings}\\
http://www.lpi.usra.edu/meetings/ppv2005/pdf/8156.pdf

\refs Claussen, M.~J., Marvel, K.~B., Wootten, A., and Wilking, B.~A.\ 
(1998) {\em Astrophys. J.}, 507, L79-L82. 

\refs Corcoran, M. and Ray, T.~P.\ (1997) {\em Astron. Astrophys., 321}, 
189-201.

\refs Corcoran, M. and Ray, T.~P.\ (1998) {\em Astron. Astrophys., 336}, 
535-538.

\refs Coffey, D., Bacciotti, F., Ray, T.~P., Woitas, J., and Eisl\"offel, J. 
(2004) {\em Astrophys. J., 604}, 758-765.

\refs Coffey, D.~A., Bacciotti, F., Woitas, J., Ray, T.~P., and 
Eisl{\"o}ffel, J.\ (2005) In {\em PPV Poster Proceedings}\\
http://www.lpi.usra.edu/meetings/ppv2005/pdf/8032.pdf

\refs Davis, C.~J., Berndsen, A., Smith, M.~D., Chrysostomou, A., and Hobson, 
J. (2000) {\em Mon. Not. Roy. Astr. Soc., 314}, 241-255.

\refs Draine, B.~T. (1995) {\em Astrophys Space Sci., 233}, 111-123.

\refs Devine, D., Grady, C.A., Kimble, R.A., Woodgate, B., Bruhweiler, F.C.,
Boggess, A., Linsky, J.L., and Clampin, M. (2000) {\em Astroph. J., 542}, 
L115-L118.

\refs Dougados, C., Cabrit, S., Ferreira, J., Pesenti, N., Garcia, P., 
and O'Brien, D. (2004) {\em Astrophys Space Sci., 292}, 643-650.

\refs Dougados, C., Cabrit, S., Lavalley, C., and M\'enard, F. (2000)
{\em Astron. Astrophys., 357}, L61-L64.

\refs Dougados, C., Cabrit, S., and Lavalley-Fouquet, C. (2002)
{\em Rev. Mex. Astron. Astrofis., 13}, 43-48. 

\refs Dougados, C., Cabrit, S., Lopez-Martin, L., Garcia, P., and 
O'Brien, D.\ (2003), {\em Astrophys. Space Sci., 287}, 135-138

\refs Dougados, C., Cabrit, S., Ferreira, J., Pesenti, N., Garcia, P., and 
O'Brien, D.\ (2004) {\em Astrophys. Space Sci., 293}, 45-52 

\refs Eisl\"{o}ffel, J., Mundt, R., Ray, T. P., and Rodr\'{\i}guez, L. F. 
(2000) In {\em Protostars and Planets IV} (V. Mannings et al., eds.), 
pp. 815-840. Univ. of Arizona, Tucson.

\refs Fern\'andez, M. and Comer\'on, F. (2001) {\em Astron. Astrophys., 380},
264-276.

\refs Fern\'andez, M., and Comer\'on, F. (2005) {\em Astron. Astrophys., 440},
1119-1126.

\refs Ferreira, J.\ (1997) {\em Astron. Astrophys., 319}, 340-359. 

\refs Ferreira, J., Dougados, C., and Cabrit, S. (2006) {\em Astron. 
Astrophys.,} in press.

\refs Finkenzeller, U. (1985) {\em Astron. Astrophys., 151}, 340-348.

\refs Folha, D.~F.~M. and Emerson, J.~P.\ (2001) {\em Astron. Astrophys.}, 
365, 90-109.

\refs Frank, A., and Mellema, G. (1996) {\em Astrophys.\ J., 472}, 684-702.
 
\refs Garcia, P.~J.~V., Thi{\'e}baut, E., and Bacon, R. (1999) 
{\em Astron. Astrophys., 346}, 892-896.

\refs Garcia,~P.J.V., Cabrit,~S., Ferreira,~J., and Binette,~L. (2001), 
{\em Astron. Astrophys., 377}, 609-616.

\refs Girart, J.~M., Curiel, S., Rodr{\'{\i}}guez, L.~F., 
and Cant{\'o}, J.\ 2002, {\em Rev. Mex. Astron. Astrofis., 38}, 
169-186. 

\refs Giannini, T., McCoey, C., Caratti o Garatti, A., Nisini, B., 
Lorenzetti, D.,
and Flower, D. R., (2004) {\em Astron. Astrophys., 419}, 999-1014.

\refs Goodson, A.~P., B{\"o}hm, K.-H., and Winglee, R.~M.\ (1999) 
{\em Astroph. J., 524}, 142-158.

\refs Gomez, M., Kenyon, S.J., and Whitney, B.A. (1997) {\em Astron. J., 114}, 
265-271.

\refs Goodrich, R. (1986) {\em Astroph. J., 311}, 882-894.

\refs Grady, C.A., Devine, D., Woodgate, B., Kimble, R., Bruhweiler, F.C.,
Boggess, A., Linsky, J.L., Plait, P., Clampin, M., and Kalas, P. (2000)
{\em Astroph. J., 544} 895-902.

\refs Grady, C.A., Woodgate, B., Torres, C.A.O., Henning, T., Apai, D.,
Rodmann, J., Wang, H., Stecklum, B., Linz, H., Willinger, G.M., Brown, A.,
Wilkinson, E., Harper, G.M., Herczeg, G.J., Danks, A., Vieira, G.L., Malumuth,
E., Collins, N.R., and Hill, R.S. (2004) {\em Astroph. J., 608}, 809-830.

\refs Hartigan, P., Edwards, S., and Gandhour, L. (1995)
{\em Astrophys. J., 452}, 736-768.

\refs Hartigan, P., Edwards, S., and Pierson, R. (2004)
{\em Astrophys. J., 609}, 261-276.

\refs Hartigan, P., Morse, J., and Raymond, J. (1994) 
{\em Astroph. J., 436}, 125-143.

\refs Hartigan, P., Raymond, J., and Hartmann, L. (1987) 
{\em Astroph. J., 316}, 323-348.

\refs Herbig, G. (1968) {\em Astroph. J., 152}, 439-441.

\refs Herbst, T. (2003) {\em Astrophy. Sp. Sci., 286}, 45-53. 

\refs Hirth, G.~A., Mundt, R., and Solf, J. (1997) {\em Astron. Astrophys. 
Suppl., 126}, 437-469.

\refs Hirth, G.~A., Mundt, R., Solf, J., and Ray, T.~P.\ (1994) 
{\em Astroph. J., 427}, L99-L102. 

\refs Jayawardhana, R., Mohanty, S., and Basri, G. (2003) {\em Astroph. J., 
592}, 282-287.

\refs K\"{o}nigl, A. and Pudritz, R. (2000)
In {\em Protostars and Planets IV} (V. Mannings et al., eds.), 
pp. 759-788. Univ. of Arizona, Tucson.

\refs Lavalley, C., Cabrit, S., Dougados, C., Ferruit, P., and 
Bacon, R.\ (1997) {\em Astron. Astrophys., 327}, 671-680.

\refs Lavalley-Fouquet, C. Cabrit, S. and Dougados, C. (2000) 
{\em Astron. Astrophys., 356}, L41-L44.

\refs Livio, M.\ (2004) {\em Baltic Astronomy, 13}, 273-279. 

\refs L{\'o}pez-Mart{\'i}n, L., Cabrit, S., and Dougados, C. (2003)
{\em Astron. Astrophys., 405}, L1-L4.
 
\refs Masciadri, E. and Raga, A.C.\ (2004) {\em Astroph. J., 615}, 850-854.

\refs Massaglia, S., Mignone, A., Bodo, G., (2005) {\em Astron. Astrophys.,
442}, 549-554

\refs Massaglia, S., Mignone, A., Bodo, G., Bacciotti, F. (2006) 
{\em Astron. Astrophys.}, in press.

\refs McGroarty, F. and Ray, T.~P.\ (2004), {\em Astron. Astrophys., 420}, 
975-986. 

\refs Mohanty, S., Jayawardhana, R., and Basri, G. (2005) {\em Astroph. J., 
626}, 498-522.

\refs Mundt, R. and Eisl{\"o}ffel, J.\ (1998) {\em Astron. J., 116}, 860-867.

\refs Muzerolle, J., Calvet, N., and Hartmann, L. (1998) 
{\em Astroph. J., 492}, 743-753.

\refs Muzerolle, J., Hillenbrand, L., Calvet, N., Brice\~no, C., and 
Hartmann, L. (2003) {\em Astroph. J., 592}, 266-281.

\refs Natta, A., Testi, L., Muzerolle, J., Randich, S., Comer\'on, F., 
and Persi, P. (2004) {\em Astron. Astrophys., 424}, 603-612.

\refs Neckel, T., Staude, H.J., Sarcander, M., and Birkle, K. 
{\em Astron. Astrophys., 175}, 231-237.

\refs Nisini, B., Bacciotti, F., Giannini, T., Massi F.,
Eisl\"{o}ffel, J., Podio, L., and Ray, T.P. (2005)
{\em Astron. Astrophys., 441}, 159-170.

\refs Nisini, B., Caratti o Garatti, A., Giannini, T., and 
Lorenzetti, D. (2002) {\em Astron. Astrophys., 393}, 1035-1051.

\refs Padoan, P., Kritsuk, A., Norman, M., and Nordlund, A. (2005) {\em
 Astroph. J., 622}, L61-L64.

\refs Pesenti, N., Dougados, C., Cabrit, S.,
O'Brien, D., Garcia, P., and Ferreira, J. (2003)
{\em Astron. Astrophys., 410}, 155-164.

\refs Pesenti, N., Dougados C., Cabrit S., Ferreira J.,
O'Brien D., and Garcia P. (2004) {\em Astron. Astrophys., 416}, L9-L12.

\refs Pyo, T.-S., et al.\ 2003, {\em Astrophy. Space Sci., 287}, 21-24.

\refs Raga, A., Cabrit, S., Dougados, C., and Lavalley, C.\ (2001) 
{\em Astron. Astrophys., 367}, 959-966

\refs Ray, T.~P., Mundt, R., Dyson, J.~E., Falle, S.~A.~E.~G., 
and Raga, A.~C.\ (1996) {\em Astrophys. J., 468}, L103-L106. 

\refs Ray, T.~P., Muxlow, T.~W.~B., Axon, D.~J., Brown, A., 
Corcoran, D., Dyson, J., and Mundt, R.\ (1997) {\em Nature, 385}, 415-417.

\refs Reid, M.~J., Argon, A.~L., Masson, C.~R., Menten, K.~M., and 
Moran, J.~M.\ (1995) {\em Astrophys. J., 443}, 238-244.

\refs Reipurth, B. and Bally, J. (2001) {\em Ann. Rev. Astron. Astrophys., 
39}, 403-455. 

\refs Reipurth, B., Bally, J., and Devine, D. (1997) {\em Astron. J., 114}, 
2708-2735.

\refs Scholz, A. and Eisl\"offel, J. (2004) {\em Astron. Astrophys., 419},
249-267.

\refs Shang, H., Glassgold, A.E., Shu, F.H., and Lizano, S. (2002)
{\em Astrophys. J., 564}, 853-876.

\refs Shu, F. H., Najita, J. R., Shang, H., and Li, Z.-Y. (2000)
In {\em Protostars and Planets IV} (V. Mannings et al., eds.), 
pp. 789-814. Univ. of Arizona, Tucson.

\refs Soker, N., (2005)  {\em Astron. Astrophys., 435}, 125-129.

\refs Solf, J. and B\"ohm, K.H. (1993) {\em Astroph. J., 410}, L31-L34.

\refs Takami, M., Bailey, J.A., and Chrysostomou, A. (2003) 
{\em Astron. Astrophys., 397}, 675-691.

\refs Takami, M., Bailey, J.A., Gledhill, T.M., Chrysostomou, A., 
and Hough, J.H. (2001) {\em Mon. Not. Roy. Astr. Soc., 323}, 177-187.

\refs Takami, M., Chrysostomou, A., Ray, T. P., Davis, C., Dent, W. R. F.,
Bailey, J., Tamura, M., and Terada, H. (2004) 
{\em Astron. Astrophys., 416}, 213-219.

\refs Testi, L., Bacciotti, F., Sargent, A. I., Ray, T. P., 
and Eisl\"offel, J. (2002), {\em Astron. Astrophys., 394}, L31-L34.

\refs Torrelles, J.~M., Patel, N., G{\'o}mez, J.~F., Anglada, G., and 
Uscanga, L. (2005), {\em Astrophys. Space Sci., 295}, 53-63. 

\refs Varni\'ere, P., Blackman, E.C, Frank, A., and Quillen, A.C. 
(2005) In {\em PPV Poster Proceedings}\\  
http://www.lpi.usra.edu/meetings/ppv2005/pdf/8064.pdf

\refs Watson, A.~M., and Stapelfeldt, K.~R. (2004)
{\em Astrophys. J., 602}, 860-874.

\refs Whelan, E., Ray, T.P., and Davis, C.J. (2004) 
{\em Astron. Astrophys., 417}, 247-261.

\refs Whelan, E.T., Ray, T.P., Bacciotti, F., Natta, A., Testi, L., and 
Randich, S. (2005) {\em Nature, 435}, 652-654.

\refs Woitas, J., Ray, T.P., Bacciotti, F., Davis, C.J., and 
Eisl\"{o}ffel, J. (2002) {\em Astrophys. J., 580}, 336-342.

\refs Woitas, J., Bacciotti, F., Ray, T.P., Marconi, A., Coffey, D., and
Eisl\"offel, J., (2005) {\em Astron. Astrophys., 432}, 149-160.

}
\end{document}